\documentclass[aps,superscriptaddress,showpacs,preprint,amsmath,amssymb]{revtex4}
\usepackage{graphicx}
\usepackage{epsfig}
\usepackage{epsf}
\usepackage{latexsym}
\usepackage{dcolumn}
\usepackage{bm}
\usepackage{subfigure}
\usepackage{comment}
\def \ee {e^+e^-}
\def \pp {\pi^+\pi^-}
\def \kk {K^+K^-}
\def \jp {J/\psi}
\def \psip {\psi(2S)}
\def \thetastar {\theta^*_\gamma}
\begin{document}
\title{An exclusive event generator for $e^+e^-$ scan experiments}
\author{Rong-Gang Ping}\affiliation{Institute of High Energy Physics,Chinese Academy of Sciences,
 P.O. Box 918(1), Beijing 100049, People's Republic of China}
\begin{abstract}
An exclusive event generator is designed for $\ee$ scan experiments, including initial state radiation
effects up to the second order correction. The generator is coded within
the framework of BesEvtGen. There are seventy hadronic decay modes available, with effective center-of-mass energy
coverage from the two pion mass threshold up to about 6 GeV. The accuracy achieved for the initial state radiation correction
reaches the level achieved by the {\sc KKMC} generator. The uncertainty associated with the
calculation of the correction factor to the initial state
radiation is dominated by the measurements of the energy-dependent Born cross section.
\end{abstract}
\maketitle
\newpage
\section{Introduction}
$\ee$ scan experiments have played an important role in the determination of the non-perturbative hadronic contribution to the running of the quantum electrodynamic (QED) fine structure constant, and to precise determination of electroweak measurements. The $R$-values, defined as the ratio of the total hadronic cross section to the $\ee\to\mu^+\mu^-$ cross section, have been measured by many collaborations either in $\ee$ scan experiments or in $B$ factories, over the effective center-of-mass energy region from the two pion threshold mass to the $\Upsilon$ peak \cite{Rvalues}. In the tau-charm energy region from 2 to 5 GeV, the $R-$values measured at BESII \cite{bes2R} were used in the 2011 evaluation of $\Delta\alpha_{\textrm{had}}^{(5)}(M_Z^2)$ and allowed
its precision to be improved by about a factor of 2 \cite{alpha}. $\ee$ experiments have also been used to study hadronic form factors, such as pion and nucleon form factors. The cross sections for a large number of
exclusive processes have been measured over the range from $m_{2\pi}$ to 5 GeV \cite{xscollection}. Due to the strong coupling of vector mesons to the $\ee$ beam via a virtual photon, the lineshape of the vector resonance can also be measured with $\ee$ scan experiments. Charmonium-like states, such as the $Y(4260)$ \cite{y4260}, $Y(4360)$ \cite{y4360} and so on, have been observed recently in $\ee$ colliders, and measurements of their decay modes, masses and widths are desirable. The $Z_c(3900)$ state in particular, observed by BESIII \cite{beszc} and then confirmed by the Belle \cite{bellezc} and CLEO \cite{cleozc} experiments, has stimulated a lot of theoretical discussion about its strange structure.

An event generator which includes initial state radiation (ISR) effects is indispensable to measure the cross section for a given exclusive decay. For low energy experiments, the ISR effects and event generators are comprehensively reviewed in Ref. \cite{isrgen}. Experimentally, the Born cross section is determined with the formula $\sigma={N_{\rm sig}\over \mathcal{L}\epsilon(1+\delta)}$, where $N_{\rm sig}$ is the number of signal events, $\mathcal{L}$ is the luminosity of the data sample, and the detection efficiency $\epsilon$ multiplied by the ISR correction factor ($1+\delta$) is provided by the event generator. Two popular event generators for this purpose are Babayaga \cite{babayaga} and PHOKHARA \cite{phokhara}, which give a high level of precision.
PHOKHARA can simulate the ISR process at next-to-leading order accuracy with nine hadronic
modes available. However, the original PHOKHARA generator is designed for the ISR return process with emission of a hard photon. The Babayaga generator is designed for
luminosity measurements with one additional hadronic mode, $\pp$, available. Event generation for hadronic decays in these two generators is based on the calculation of decay amplitudes with the input of hadron form factors.

Based on the requirements of $\ee$ scan experiments in the tau-charm region, we design an exclusive event generator with the ISR effects up to the second order correction included. The accuracy achieved for the ISR calculation
can reach the level achieved by the KKMC generator. The total cross section is calculated using the experimental
Born cross sections convoluted with the radiator function, rather than by
the calculation of the decay amplitude. This method
facilitates the use of full experimental information, and avoids being constrained by the form factor calculation. This method is similar to that adopted in the measurement of charm production cross section in the $\ee$ scan experiment \cite{ddxs}.

\section{Calculation of production cross section}
\begin{figure}[htbp]
\begin{center}
\epsfysize=4cm \epsffile{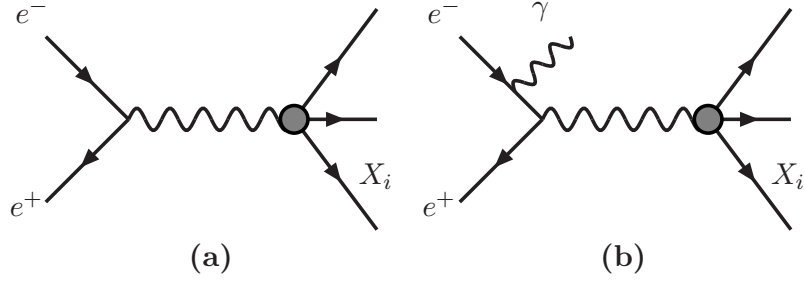}
\caption{Feynman diagrams for the process (a) $\ee\to X_i$, and ISR process (b) $\ee\to\gamma_{\rm ISR} X_i$  \label{fig1}.}
\end{center}
\end{figure}

In $\ee$ scan experiments, we consider measuring the Born cross section ($\sigma_0$) for an exclusive process
$\ee\to X_i$, as shown in Fig. \ref{fig1} (a), where $X_i$ denotes the hadron final states. Due to the ISR, what the experiments directly measure is the observed cross section ($\sigma$) for the process $\ee\to\gamma_{\rm ISR}X_i$, as shown in Fig. \ref{fig1} (b). The observed cross section is related to the Born cross section by the quasi-real electron method \cite{bornXS}:
\begin{equation}
{d^2\sigma(s)\over dmd\cos\theta_\gamma^*}={2m\over s}W(s,x,\theta^*_\gamma)\sigma_0(m),
\end{equation}
where $m$ is the invariant mass of the final states; $s$ is the full $\ee$ center-of-mass energy squared;  $x\equiv2E^*_\gamma/\sqrt s=1-m^2/s$, and $E^*_\gamma$ and $\theta^*_\gamma$ are the ISR photon energy and polar angle, respectively, in the $\ee$ center-of-mass frame. At the first order of QED calculation, the radiative function takes the form $W(s,x,\theta^*_\gamma)$ \cite{radfunction},
\begin{equation}\label{angISR}
W(s,x,\thetastar)={\alpha\over \pi x}\left( {2-2x+x^2\over \sin^2\thetastar}-{x^2\over 2}\right),
\end{equation}
which is the probability of ISR photon emission for $\thetastar\gg m_e/\sqrt s$, where $\alpha$ is the fine-structure constant and $m_e$ is the electron mass. If one performs integration over $\theta^*_\gamma$, the following energy dependence is obtained:
\begin{equation}{\label{radxs}}
\sigma(s)=\int_{M_{\rm{th}}}^{\sqrt s} dm {2m\over s}W(s,x)\sigma_0(m),
\end{equation}
where the radiative function takes $W(s,x)={2\alpha\over \pi x}(L-1)(1-x+{x^2\over 2})$ with $L=2\ln{\sqrt s\over m_e}$, $M_{\rm{th}}$ is the energy threshold value of a given decay, and $x\equiv1-m^2/s$. Note that, to the first order accuracy of QED calculation, the cross section given in Eq. (\ref{radxs}) is infrared divergent. It is therefore necessary to go beyond the first order calculation to sum all diagrams with soft multi-photon emission; this procedure leads to soft-photon exponentiation. To calculate the finite-order leading logarithmic correction, the structure function method is suggested in Ref. \cite{fadin}. This method yields the same factorized form (\ref{radxs}) for the radiative photon emission cross section. Up to order $\alpha^2$, the radiative function takes the form:
\begin{equation}{\label{secondRad}}
W(s,x)=\Delta\cdot\beta x^{\beta-1}-{\beta\over 2}(2-x) + {\beta^2\over 8}\{(2-x)[3\ln(1-x)-4\ln x]-4{\ln(1-x)\over x}-6+x\},
\end{equation}
where
\begin{eqnarray}
L&=&2\ln {\sqrt s \over m_e},\\
\Delta&=&1+{\alpha\over\pi}({3\over 2}L+{1\over 3}\pi^2-2)+({\alpha\over \pi})^2\delta_2,\\
\delta_2&=&({9\over 8}-2\xi_2)L^2-({45\over 16}-{11\over 2}\xi_2-3\xi_3)L-{6\over 5}\xi_2^2-{9\over 2}\xi_3-6\xi_2\ln 2+{3\over 8}\xi_2+{57\over 12},\nonumber\\
\beta&=&{2\alpha\over \pi}(L-1),~\xi_2=1.64493407,~\xi_3=1.2020569.
\end{eqnarray}
Here the exponentiated part in Eq. (\ref{secondRad}) accounts for soft multi-photon
emission, while the remaining part takes into account hard collinear bremsstrahlung
in the leading logarithmic approximation. We use the radiative function up to second
order calculation to determine the cross section;
 its accuracy is high enough for us to construct the event generator, though contributions from the $\alpha^2$-order
  are known \cite{thirdRad}.
\section{Event generator}
The event generator is constructed using the observed cross section as given in Eq. (\ref{radxs}),
 and the integral is decomposed into two parts:
\begin{equation}
\sigma(s)\equiv\sigma^I(s)+\sigma^{II}(s)=\int_{M_{\rm th}}^{M_0} dm {2m\over s}W(s,x)\sigma_0(m)
+\int_{M_{0}}^{\sqrt s} dm {2m\over s}W(s,x)\sigma_0(m),
\end{equation}
where the threshold energy $M_{\rm th}$ is the sum of masses for the final states,
and the integral is cut apart at the point $M_0=\sqrt{s-2\sqrt sE_\gamma^{\rm cut}}$ with an energy cut $E^{\rm cut}_\gamma$ on the ISR photon. In practice, $E_\gamma^{\rm cut}$ is set to the energy sensitivity of photon detection, for example, 25 MeV for the BESIII detector. It means that an ISR photon within the range $0\sim E_\gamma^{\rm cut}$ is too soft to be detected, so events within this energy region are generated with the ISR photon suppressed. To simplify the calculation, Born cross sections near the energy point $\sqrt s$ are assumed to be a constant value $\sigma_0(\sqrt s)$. Using the relation $x=1-m^2/s$, the second integral can be further decomposed into two parts:
\begin{equation}
\sigma^{II}(s)=\int_{M_{0}}^{\sqrt {s(1-b)}} dm {2m\over s}W(s,x)\sigma_0(m)+\sigma_0(\sqrt s)\lim_{a\to0}\int_a^bW(s,x)dx,
\end{equation}
with $b\ll1$. Using the radiative function given in Eq. (\ref{secondRad}), one has
\begin{eqnarray}
\lim_{a\to0}\int_a^bW(s,x)dx&=&\Delta b^\beta+{\beta^2b^2\over 32}+{\beta b^2\over 4}-{3\over16}\beta^2b^2\ln (1-b)
+{1\over 4}\beta^2 b^2\ln b-{5\over 16}\beta^2 b
-\beta b\nonumber\\
&&+{3\over 4}\beta^2b\ln(1-b)-\beta^2b\ln b-{9\over 16}\beta^2\ln(1-b)+{1\over 2}\beta^2\textrm{Li}_2(b),
\end{eqnarray}
with Spence's function $\textrm{Li}_2(x)=-x+{1\over 4}x^2-{1\over 9}x^3~(x\ll1)$.

The Monte Carlo (MC) technique of acceptance-rejection method \cite{pdg} is used to generate events. First, an event with ($N^I$) or without ($N^{II}$) ISR photons is sampled with the cross section $\sigma^I(s)$ and $\sigma^{II}(s)$. A random number $r_1\in(0,1)$ is generated; if the condition $r_1<{\sigma^I(s)\over \sigma^I(s)+\sigma^{II}(s)}$ is true, then the generator produces an event of type $N^{I}$, otherwise, produces an $N^{II}$-event. The decays of event type $N^{I}$ proceed via a further three steps:
\begin{itemize}
\item[(1)] An unweighted event of the decay $\ee\to\gamma_{\rm ISR} X_i$ is generated within the uniform phase space with an invariant
hadron mass $M_{\rm hds}$.
\item[(2)] A random number $r_2\in(0,1)$ is then generated, and tested to see whether $r_2\le{\sigma^I(M_{\rm hds}^2)\over \sigma^I_{max}(s)}$ is true. If so, then the event is accepted; otherwise, the event is rejected and the program goes back to step one. Here $ \sigma^I_{\rm max}(s)$ is the maximum value of the observed cross section within the energy range from
$M_{\rm th}$ to $M_0$.
\item[(3)]The last step is to sample the ISR photon polar angle according to Eq. (\ref{angISR}). A random number $r_3\in(0,1)$ is generated and tested to check if $r_3\le{W(s,x,\theta^*_\gamma)\over W_{\rm max}(s,x)}$ is true; if so, the event is accepted, and otherwise the event is rejected and the program goes back to step one. To avoid collinear divergence when $\theta_\gamma^*=0$, it is required that $1^\circ\le\theta_\gamma^*\le189^\circ$, which is large enough to cover the BESIII detector barrel and endcap. Here $W_{\rm max}(s,x)$ is the maximum value of $W(s,x,\theta^*_\gamma)$ over this ISR photon polar angle region.
\end{itemize}

Currently the generator includes seventy exclusive decays in total, with energy region covering a range from 0.3 GeV up to about 6 GeV. Their Born cross sections are collected from the published reference papers. They are: $\ee\to$ $p\bar p$ \cite{babar_ppbar}, $n\bar n$ \cite{infn}, $\Lambda\bar \Lambda$ \cite{babar_llbar},
$\Sigma\bar \Sigma^0$ \cite{babar_llbar}, $\Lambda\bar \Sigma^0$ \cite{babar_llbar}, $\Sigma^0\bar \Lambda $
\cite{babar_llbar}, $\pi^+\pi^- $ \cite{babar_pipi}, $\pi^+\pi^-\pi^0 $
\cite{barbar_3pi}, $K^+K^-\pi^0 $ \cite{babar_kkpi0}, $K_SK^+\pi^- $ \cite{babar_kkpi0},
$K_SK^-\pi^+$ \cite{babar_kkpi0}, $K^+K^-\eta $ \cite{babar_kkpi0}, $2(\pp) $
\cite{babar_4pi}, $\pp2\pi^0$ \cite{babar_pipi2pi0}, $\kk\pp $
\cite{babar_kkpp}, $\kk 2\pi^0 $ \cite{babar_kkpp}, $2(\kk) $ \cite{babar_4k}, $2(\pp)\pi^0$
\cite{babar_5pi}, $2(\pp)\eta $ \cite{babar_5pi}, $\kk\pp\pi^0 $ \cite{babar_5pi}, $\kk\pp\eta $ \cite{babar_5pi},
$3(\pp) $ \cite{babar_6pi}, $2(\pp\pi^0) $ \cite{babar_6pi}, $\phi\eta $ \cite{babar_kkpi0}, $\phi\pi^0 $\cite{babar_kkpi0},
 $K^+K^{*-} $ \cite{babar_kkpi0},$K^-K^{*+} $\cite{babar_kkpi0}, $K_S\bar K^{*0}(892) $ \cite{babar_kkpi0},
  $K^{*}(892)^0K^+\pi^- $ \cite{babar_kkpp}, $K^{*}(892)^0K^-\pi^+$ \cite{babar_kkpp},
 $K^{*}(892)^-K^+\pi^0 $ \cite{babar_kkpp}, $K^{*}(892)^+K^-\pi^0 $ \cite{babar_kkpp},
$K_2^*(1430)^0K^+\pi^- $ \cite{babar_kkpp},
$K_2^*(1430)^0K^-\pi^+ $ \cite{babar_kkpp},
 $\kk\rho $ \cite{babar_kkpp},
 $\phi\pp $ \cite{babar_kkpp},
$\phi f_0(980) $, \cite{babar_kkpp},
 $\eta\pp $  \cite{babar_llbar},
$\omega\pp $  \cite{babar_llbar},
 $\omega f_0(980) $ \cite{babar_llbar},
 $\eta'\pp $  \cite{babar_llbar},
 $f_1(1285)\pp $  \cite{babar_llbar},
 $\omega\kk $  \cite{babar_llbar},
 $\omega\pp\pi^0 $  \cite{babar_4pi},
 $\Sigma^-\bar \Sigma^+ $ \cite{babar_llbar},
$K^+K^-$  \cite{kkbar,ppkkpipi},
 $K_{S}K_{L}$  \cite{kkbar},
 $p\bar p\pi^0$,
$p\bar p\eta$,
 $D^0\bar D^{*0}$ \cite{ddbar},
 $\bar D^0 D^{*0}$\cite{ddbar},
$D^0\bar D^0$\cite{ddbar},
 $D^+D^-$ \cite{ddbar},
 $D^+D^{*-}$ \cite{ddstar},
 $D^-D^{*+}$ \cite{ddstar},
 $D^{*+}D^{*-}$ \cite{ddstar},
 $D^0D^-\pi^+$\cite{ddpi},
 $\bar D^0D^+\pi^-$\cite{ddpi},
 $D^0D^{*-}\pi^+$\cite{ddstarpi},
 $\bar D^0D^{*+}\pi^-$\cite{ddstarpi},
 $\Lambda_c^+\Lambda_c^-$\cite{lambdacpair},
$\eta\jp$ \cite{etajsi},
 $\jp\pp$ \cite{pipijsi},
 $\pi^+\pi^-h_c$,
 $\pi^0\pi^0h_c$,
 $\psip\pp$ \cite{pipipsip},
 $\jp\kk$ \cite{kkjsi},
 $D_s^+D_s^-$ \cite{dsds},
 $D_s^{*+}D_s^-$ \cite{dsds},
 $D_s^{*-}D_s^+$ \cite{dsds}.

\begin{figure}
 \centering
 \begin{tabular}{cc}
  \epsfig{file=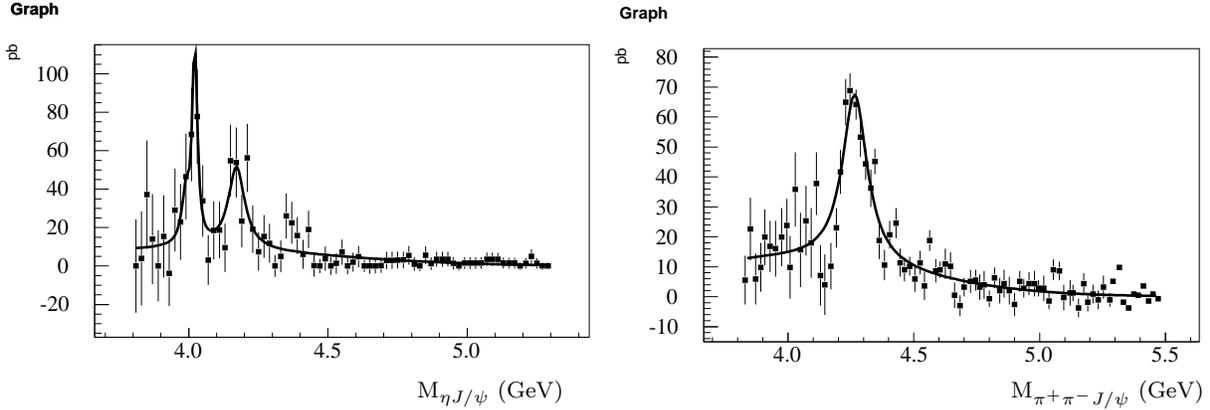,width=0.95\linewidth,clip=}
  \end{tabular}
  \caption{The cross sections for $\ee\to\eta\jp$ (left) and $\pp\jp$ (right). The points with error bars are the measurements;
  the curves are fitted results parameterized with a multi-Gaussian function.
  }
  \label{xsdis}
\end{figure}

The function for the Born cross section $\sigma_0(s)$ is parameterized as a multi-Gaussian function; its parameters are determined by fitting the cross section mode by mode, requiring the fit quality to be acceptable. As an example, Fig. \ref{xsdis} shows the cross sections from experimental measurements for $\ee\to\eta\jp$ and $\pp\jp$. The fitted values shown in the curves are used as the input to the Born cross section function in the generator.

\begin{figure}
 \centering
 \begin{tabular}{cc}
  \epsfig{file=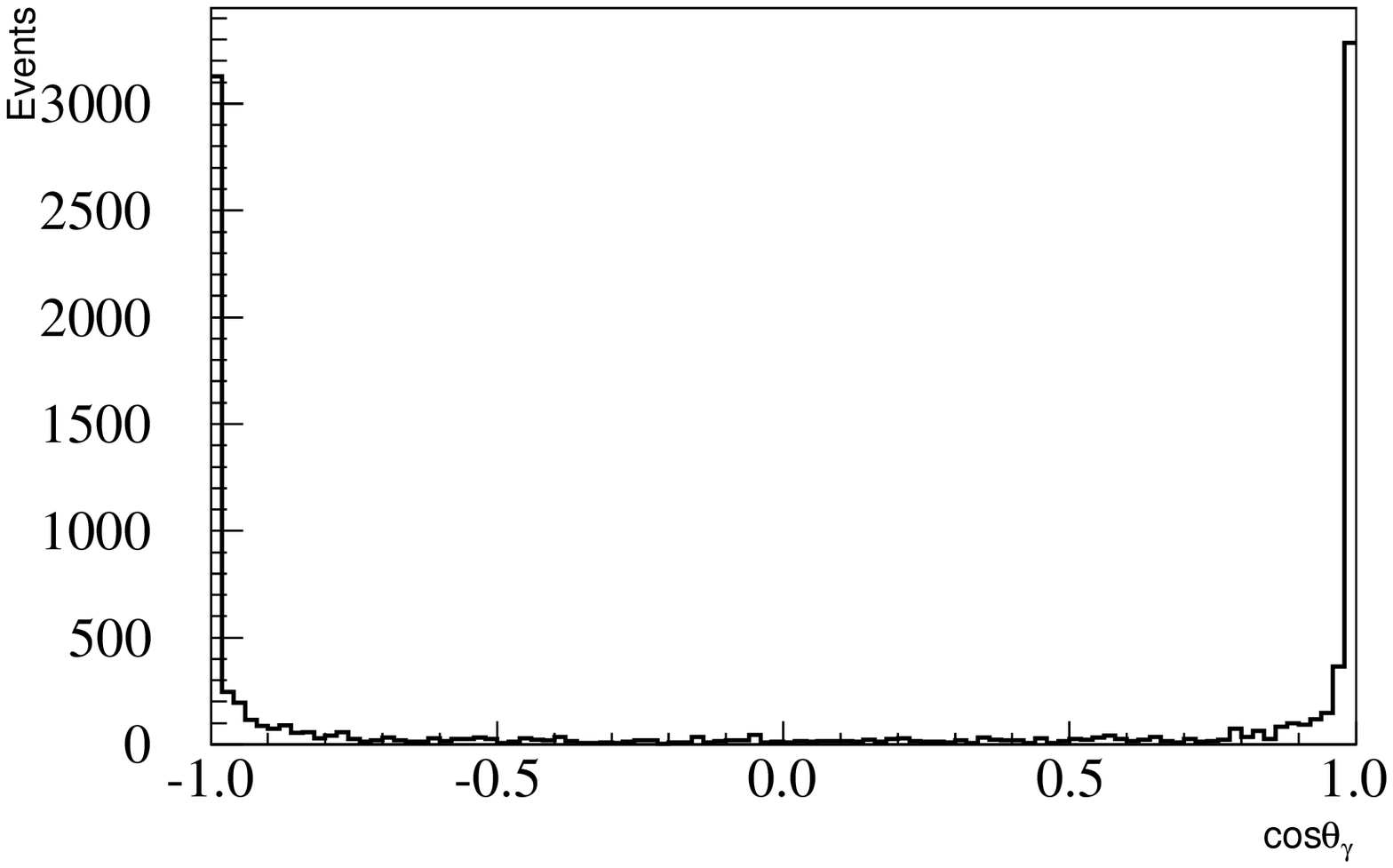,width=0.45\linewidth,clip=} &
  \epsfig{file=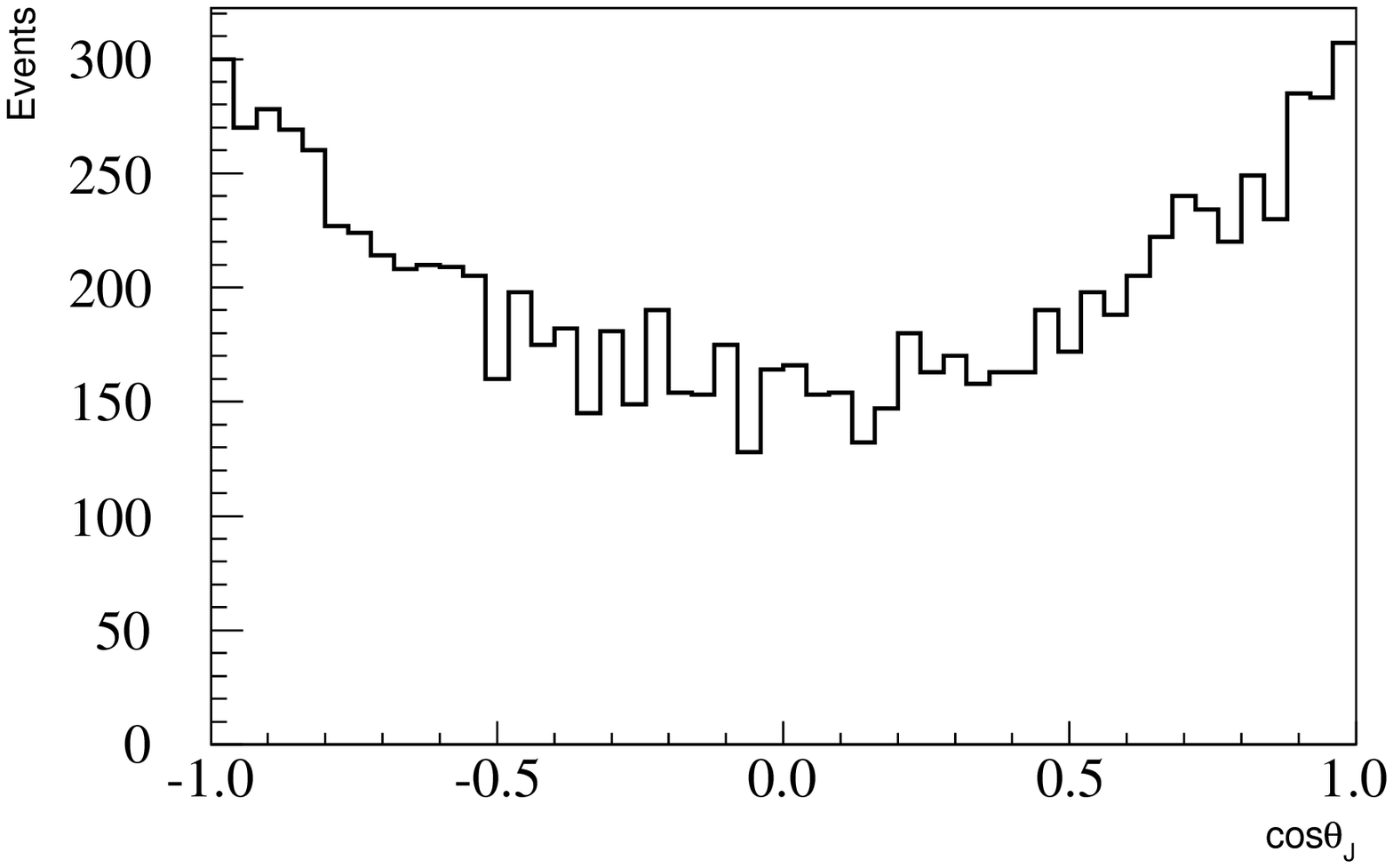,width=0.45\linewidth,clip=}
  \end{tabular}
  \caption{The generated distributions for the process $\ee\to\eta\jp$. Left panel: the ISR photon
  polar angle distribution. Right panel: the $J/\psi$ polar angle distribution.}
  \label{MC}
\end{figure}

The angular distribution for ISR photons is implemented according to Eq. (\ref{angISR}), characterized by the beam collinear distribution, as shown in Fig. \ref{MC}. Angular distributions for final states, however, are implemented only for the two-body decay, namely, $1+\cos^2\theta$ for PP and BB(where P is a pseudoscalar meson, B is a baryon) modes, and $1-\cos^2\theta$ for VP (where V is a vector meson) modes. Unfortunately, the information on the intermediate states and the angular distribution for the multibody decays are not available for the above processes. If this information becomes available from experiments in the future, they will be implemented in the generator. Figure \ref{MC} (right plot) shows the generated angular distribution of $\jp$ for the process $\ee\to\eta\jp$.

\section{Discussion and summary}
The generator is coded in the framework of BesEvtGen \cite{besevtgen}. To validate the program,  we compare the calculated cross section with that in the generator {\sc PHOKHARA}. The energy cut for the emission photon is set as $E_\gamma>0.025$ GeV in both generators. Table \ref{xscomp} lists the computed cross sections of $\ee\to\pp,~\pp\pi^0,~\pp2\pi^0,~2(\pp)$ and $p\bar p$. Here, the errors in our generator are due to the uncertainties of the measured cross sections, while the errors in {\sc PHOKHARA} are statistical. The calculated cross sections are consistent with each other within 2-3 $\sigma$ accuracy.

\begin{table}
\caption{Comparison of ISR cross section calculated in our generator and in {\sc PHOKHARA} with
the ISR photon energy cut $E_{\gamma}>0.025$GeV. \label{xscomp}}
\begin{tabular}{cccc}
\hline\hline
Mode & hadron mass (GeV)& Our generator & PHOKHARA\\\hline
$\pi^+\pi^-$ & $M_{2\pi}\sim3.65$ & $1.29\pm0.02$ nb & $1.26\pm0.01$ nb\\
$\pi^+\pi^-\pi^0$ & $1.06\sim 2.9$ & $89.7\pm2.7$ pb & $91.2\pm0.00$ pb \\
$\pp2\pi^0$&$M_{4\pi}\sim2.5$&$1.46\pm0.07$ nb&$1.29\pm0.01$ nb\\
$2(\pp)$&$M_{4\pi}\sim3.0$&$0.55\pm0.04$ nb&$0.49\pm0.01$ nb\\
$p\bar p$ &  $M_{p\bar p}\sim3.0$  & $18.2\pm3.3$ pb & $19.9\pm0.0$ pb\\
\hline\hline
\end{tabular}
\end{table}

To estimate the accuracy of the ISR correction, we compute the correction factors for the process $\ee\to\pp\jp$,
using KKMC for comparison. In KKMC the ISR correction to the process $\ee\to f\bar f$ (where $f$ is a fermion)
is calculated with the so-called CEEX or EEX amplitude technique~\cite{kkmc} with multi-photon emission.
The Born cross section for $f\bar f$ production in KKMC is replaced with that for the process $\ee\to\pp\jp$ as shown in Fig. \ref{xsdis}. The ISR correction factor is defined as $1+\delta=\sigma(s)/ \sigma_0(s)$, where $\sigma(s)$ is the total cross section due to the ISR correction, $\sigma_0(s)$ is the Born cross section, and $\sqrt s$ is the centre-of-mass energy of the $\ee$ beam. Table \ref{isrcmp} lists the ISR correction factor calculated with our generator and with KKMC. They are in good agreement, with differences of less than 0.5\%.
We conclude that the accuracy for the ISR correction calculation in our generator is achieved with the same level as KKMC, about 0.1-0.4\%\cite{kkmcPrecision}. The uncertainty in our generator is dominated by the uncertainty of the measured Born cross section in experiments.

\begin{table}
\caption{Comparison of ISR correction factors calculated in our models and the KKMC generator. \label{isrcmp}}
\begin{tabular}{c|cccccccc}
\hline\hline
$\sqrt s$ GeV & 4.19 & 4.21& 4.22& 4.23& 4.245& 4.26 &4.31& 4.36 \\\hline
KKMC&0.830&0.815&0.810&0.808&0.808&0.817&0.916&1.040 \\
Our generator &0.826&0.813&0.808&0.808&0.806&0.816&0.916&1.038\\
\hline\hline
\end{tabular}
\end{table}

For the events generated at a given energy point $\sqrt s$, the larger factor $(1+\delta)$ implies that there are more events returning to the low mass energy region by the emission of a photon, thus leading to a lower detection efficiency $\epsilon$ if the conservation of total energy and momentum is required in the event selection. Thus one can prove that the calculation of the factor product $\epsilon(1+\delta)$ is irrelevant to the Born cross
section below the energy point where the ISR photon can survive the event selection criteria.
Experimentally, the requirement on the kinematic fit to the final states will
veto events with an energetic ISR photon, {\it i.e.} a few hundred MeV, in the event selection. As per this argument, one concludes that the coverage of the Born cross section in this model is wide enough to get the factor $\epsilon(1+\delta)$.

The multibody decays are generated without considering the possible intermediate states and angular distributions in the generator. These can be further implemented within the framework of BesEvtGen, if the information is available in the data (as explained in the manual \cite{manual}). For the decay $\ee\to B\bar B$, the angular distribution may deviate from the form ${dN\over d\cos\theta}\propto (1+\cos^2\theta)$, where $\theta$ is the polar angle of baryon, especially near the $B\bar B$ mass threshold. If information about the baryon angular distribution is available from experiment, it can be included in the model. The final state radiation effects are not considered in this generator, but they could be implemented within the BesEvtGen framework using the PHOTOS package \cite{photos}.

To summarize, we have designed an exclusive event generator for $\ee$ scan experiments with the initial state radiation effects up to the second order correction included. The total cross section is calculated
using the Born cross sections measured in experiments as input. There are seventy hadronic decay modes available, with energy coverage from the threshold of the two pion mass up to about 6 GeV. The accuracy achieved for the initial state radiation calculation reaches the level achieved by the KKMC generator.
The uncertainty associated with the calculation of the ISR correction factor
is dominated by the errors in the energy-dependent Born cross section measured in experiments.

\section*{Acknowledgments}
I would like to thank Dr. Kai Zhu and Prof. Chenping Shen for refining the text, and I am especially grateful to Ruiling Yang for helping me to collect the experimental data.
This work is partly supported
by the National Natural Science Foundation of China under Grant Nos.
11079038, 11235011, 11375205.

\end{document}